\documentclass[pra,twocolumn,amsmath,amssymb,superscriptaddress,floatfix]{revtex4-1}

\usepackage{graphicx}
\usepackage{dcolumn}
\usepackage{bm}
\usepackage[sort&compress]{natbib}
\usepackage{xspace}
\usepackage{amsmath}
\usepackage{epstopdf}
\usepackage{amssymb}
\usepackage{epsfig}
\usepackage{textcomp}
\usepackage{changepage}

\newcommand{\mum}{\text{\textmu m}}
\newcommand{\SiN}{\text{Si}_3\text{N}_4}

\begin{document}

\title{Slot-Mode Optomechanical Crystals: A Versatile Platform for Multimode Optomechanics}

\author{Karen E. Grutter} \email{karen.grutter@nist.gov}
\affiliation{Center for Nanoscale Science and Technology, National Institute of Standards and Technology, Gaithersburg, MD 20899-6203 USA}
\author{Marcelo I. Davan\c{c}o}
\affiliation{Center for Nanoscale Science and Technology, National Institute of Standards and Technology, Gaithersburg, MD 20899-6203 USA}
\author{Kartik Srinivasan} \email{kartik.srinivasan@nist.gov}
\affiliation{Center for Nanoscale Science and Technology, National Institute of Standards and Technology, Gaithersburg, MD 20899-6203 USA}

\date{\today}

\begin{abstract}
We demonstrate slot-mode optomechanical crystals, a class of device in which photonic and phononic crystal nanobeam resonators separated by a narrow slot are coupled through optomechanical interactions. In these geometries, nanobeam pairs are patterned so that a mechanical breathing mode is confined at the center of one beam, and a high quality factor ($Q_o\gtrsim10^5$) optical mode is confined in the slot between the beams. Here, we produce slot-mode devices in a stoichiometric $\SiN$ platform, with optical modes in the 980~nm band, coupled to breathing mechanical modes at 3.4~GHz, 1.8~GHz, and 400~MHz. We exploit the high $\SiN$ tensile stress to achieve slot widths down to 24~nm, which leads to enhanced optomechanical coupling, sufficient for the observation of optomechanical self-oscillations at all studied frequencies.  We utilize the slot mode concept to develop multimode optomechanical systems with triple-beam geometries, in which two optical modes are coupled to a single mechanical mode, and two mechanical modes are coupled to a single optical mode. This concept allows great flexibility in the design of multimode chip-scale optomechanical systems with large optomechanical coupling at a wide range of mechanical frequencies.
\end{abstract}


\maketitle

\section{Introduction}

Sideband-resolved cavity optomechanical systems have recently demonstrated their potential in a wide variety of applications, including motion sensing~\cite{ref:Schliesser_resolved_sideband2, ref:Teufel_nanomechanical_motion_Heisenberg}, ground state cooling~\cite{ref:Teufel_ground_state, ref:Chan_Painter_ground_state}, and optomechanically-induced transparency~\cite{ref:Weis_Kippenberg, ref:safavi-naeini4}. For these applications, high efficiency requires large optomechanical coupling strength in addition to sideband resolution (mechanical frequency $\gg$ optical linewidth).  Additional phenomena have been observed in multimode cavity optomechanical systems, in which multiple optical and/or mechanical modes interact, including wavelength conversion~\cite{ref:Hill_Painter_WLC_Nat_Comm, ref:Liu_yuxiang_wlc, ref:Dong_Wang_Science_dark_mode}, Raman-ratio thermometry~\cite{purdy2014optomechanical}, energy transfer between mechanical modes~\cite{shkarin2014optically}, and optomechanical mode mixing~\cite{massel2012multimode}.  Phonon pair generation~\cite{dong2014optomechanically}, mechanical mode entanglement~\cite{massel2012multimode, wang2013reservoir}, and unresolved sideband cooling~\cite{ojanen2014ground} have also been theoretically proposed. In all these systems, improved performance and broader applicability could be achieved if the optical and mechanical modes could be independently tailored to a given application.

\begin{figure*}[htbp]
\centering
\includegraphics[width=0.8\linewidth]{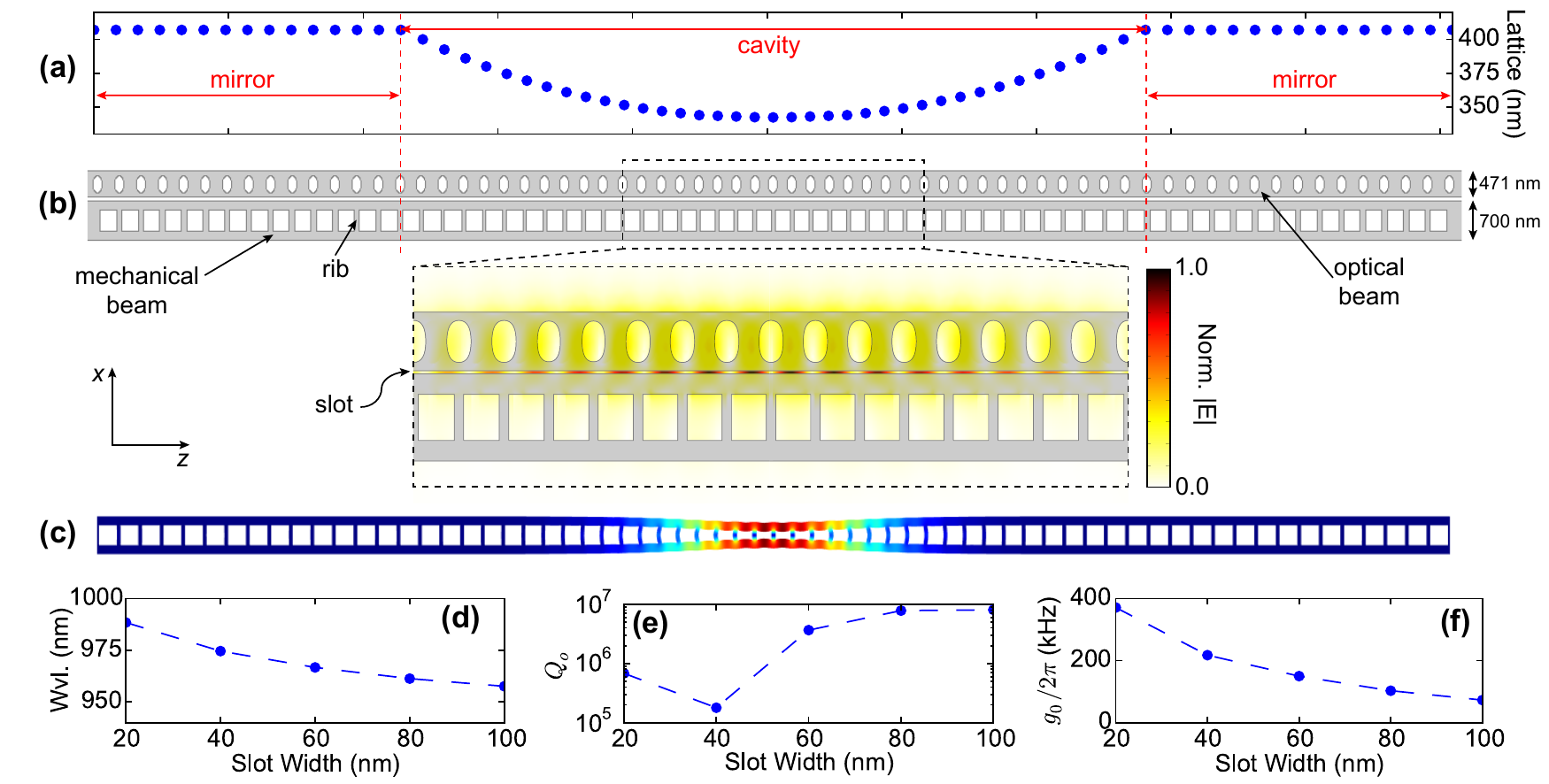}
\caption{(a) Variation of the optomechanical crystal lattice constant along the length of the beams. The period is fixed in the mirror regions at the beam ends and varies quadratically in the center cavity region.  (b) The slot mode optomechanical crystal is formed by parallel optical and mechanical beams that are separated by a narrow slot.  The zoomed-in image of the center shows the finite element method (FEM) simulated electric field amplitude of the optical slot mode around 980~nm.  (c) FEM simulation of the breathing mode of the mechanical beam (around 3.4~GHz).  (d), (e), and (f): The width of the slot is varied in an FEM simulation of the (d) resonant wavelength, (e) optical quality factor ($Q_o$), and (f) optomechanical coupling $g_0/(2\pi)$.}
\label{fig:Schem}
\end{figure*}

The slot-mode optomechanical crystal structure, in which the optical and mechanical modes are confined in separate but interacting beams (Fig.~\ref{fig:Schem}), is one method of achieving this flexibility while maintaining large optomechanical coupling strength.  Simulations~\cite{ref:Davanco_OMC} have shown that, in systems in which the optomechanical interaction is dominated by moving boundaries, this geometry can significantly increase the optomechanical coupling strength relative to single nanobeam optomechanical crystals.  It also provides the desired design flexibility to enable multimode applications such as optomechanical wavelength conversion.

In this work, we experimentally demonstrate slot-mode optomechanical crystals implemented in stoichiometric $\SiN$, a material whose broad optical transparency and large intrinsic tensile stress make it attractive for many applications.  In Sec.~\ref{sec:fab}, we show how this intrinsic stress can be exploited to achieve slots with aspect ratios of 10:1, and in Sec.~\ref{sec:initmeas} we demonstrate how tuning this slot width improves device performance in 3.4~GHz band devices.  Sec.~\ref{sec:LFMF} shows how the mechanical mode frequency can be changed while minimally affecting the optical mode, with demonstrations of 1.8~GHz and 400~MHz band devices.  Finally, in Sec.~\ref{sec:TNBs}, we extend the slot-mode optomechanical crystal concept to multimode optomechanical devices, in which two mechanical modes are coupled to a single optical mode, and two optical modes are coupled to a single mechanical mode.

\section{Basic Device Design}
\label{sec:design}

A slot-mode optomechanical crystal, shown in Fig.~\ref{fig:Schem}b, consists of two parallel beams separated by a narrow slot.  The ``optical beam'' is a photonic crystal cavity designed to confine the optical mode in the slot.  The ``mechanical beam'' is a phononic crystal resonator optimized to confine the mechanical breathing mode (Fig.~\ref{fig:Schem}c) while maintaining low optical loss.  Both the optical and mechanical modes are confined along the $z$-axis by periodic patterning of holes.  In the outer mirror region, the lattice spacing is constant, but it varies quadratically in the cavity region (Fig.~\ref{fig:Schem}a).  Details on the design of this device are outlined in Ref.~\cite{ref:Davanco_OMC}.  The devices in this work were designed for optical modes around 980~nm and mechanical breathing modes around 3.4~GHz.  The optical beam is patterned with identical elliptical holes (188~nm$\times$330~nm) along its length, while the mechanical beam holes have a constant height (370~nm) and widths that are varied such that the ``ribs'' between the holes align with the elliptical holes in the optical beam.

There have been several demonstrations of sideband-resolved single-nanobeam optomechanical crystals~\cite{ref:Chan_Painter_ground_state,ref:safavi-naeini4, ref:Hill_Painter_WLC_Nat_Comm,ref:Davanco_nanobeam_OMC, ref:Bochmann_Cleland_AlN_uwave_optomechanics}, in which a GHz frequency mechanical breathing mode is coupled to an optical mode localized by the same physical structure. These geometries are distinguished by the breathing mode's high frequency (enabling sideband resolution), isolation from mechanical supports due to the phononic mirrors, and strong interaction with the optical mode.  Our goal in this work is to retain these advantageous features while increasing the system's versatility through the slot mode geometry.

Optical slot modes have been utilized before to achieve large optomechanical coupling in microrings/disks~\cite{ref:Wiederhecker_Lipson, ref:Lin5}, bilayer photonic crystal slabs~\cite{ref:Roh}, and photonic crystal zipper cavities~\cite{ref:eichenfield1}.  These applications were lower frequency ($< 150$~MHz) than the 3.4~GHz band breathing modes in this work, and, thus operated in the unresolved-sideband regime (mechanical frequency $<$ optical linewidth).  In addition, previous demonstrations have not taken full advantage of the flexibility of the slot mode architecture, as the mechanical and optical modes were supported by the same structural components.  Separating the optical and mechanical modes into two beams enables independent design of these modes.  This opens a wide range of frequency combinations that would be difficult to access with a single optomechanical structure.  The slot-mode structure also opens the possibility for additional interactions with both modes, which can be separately accessed from the beam sides opposite the slot.  For example, electrodes could be added to the outside of the mechanical beam with minimal perturbation of the optical mode.  Adding more optical or mechanical beams, thereby forming more slots, can also increase the device functionality by realizing multimode optomechanical systems, as discussed in Secs.~\ref{sec:TNBs} and \ref{sec:Disc}.

\section{Stress Tuning and Device Fabrication}
\label{sec:fab}
\begin{figure*}[htbp]
\centering
\includegraphics[width=0.9\linewidth]{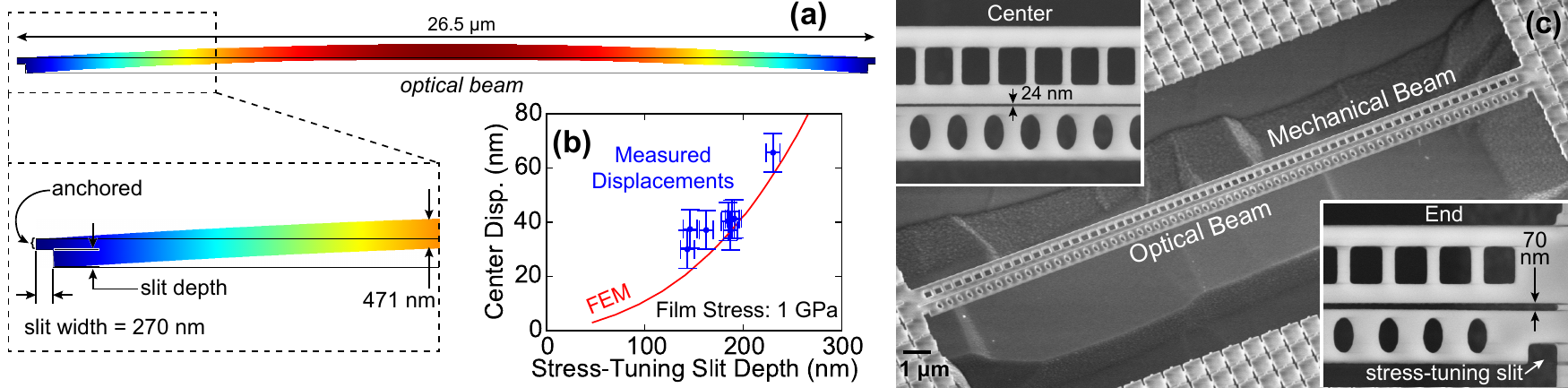}
\caption{(a) FEM simulation of a tensile-stressed beam with stress-tuning slits at the ends.  (b) Displacement at beam center with respect to slit depth.  FEM results (line) are for a beam with the same dimensions as the optical beam of the slot-mode device. Error bars on the measured data are due to the uncertainty in the SEM measurements and are one standard deviation values. (c) Scanning electron microscope (SEM) images of a released device.  Insets show the slot width at the beam end is about 70~nm, shrinking to 24~nm at the beam center.}
\label{fig:StressAndFab}
\end{figure*}
In addition to the design of the mechanical and optical beams, device parameters are also strongly dependent on the width of the slot between the two beams, simulated in Fig.~\ref{fig:Schem}d-f.  Given a device with fixed design of the optical and mechanical beams, reducing the slot width redshifts the optical resonance, and reduces $Q_o$ somewhat (still above $10^5$). The optomechanical coupling rate $g_0$ increases significantly as the slot width decreases, so the slot between the optical and mechanical beams should be made as small as possible.  Lithographically defining small spaces and etching high-aspect-ratio trenches are both challenging in fabrication.  This can be mitigated by taking advantage of the intrinsic film stress of stoichiometric $\SiN$ ($\approx 1$~GPa).  An asymmetric anchoring condition in a doubly-clamped beam induces asymmetry in the stress, thereby causing it to move laterally.  Long, thin tethers asymmetrically attached to the ends of parallel nanobeams have been used to shrink gaps to as small as 40~nm after release \cite{ref:camacho}.  To achieve the same effect, we investigated small slits at a beam's ends (Fig.~\ref{fig:StressAndFab}a).  Finite element method (FEM) simulations show that varying the width and depth of these slits controls the lateral displacement of the center of the beam (Fig.~\ref{fig:StressAndFab}b).  In the slot-mode device, a large initially defined and etched slot would be reduced post-release to the desired width by including these stress-tuning slits at the ends of the optical beam.

Slot mode optomechanical crystal nanobeams were fabricated in 250~nm thick stoichiometric $\SiN$ deposited via low-pressure chemical-vapor deposition on a bare Si substrate (tensile stress $\approx 1$~GPa).  Devices were patterned via electron-beam (E-beam) lithography in a positive E-beam resist and developed in hexyl acetate at 7~$^\circ$C.  The pattern was transferred to the $\SiN$ using a CF$_4$/CHF$_3$ reactive ion etch.  Devices were released in a 45~\% KOH solution at 75~$^\circ$C followed by a dip in a 1:4 HCl:H$_2$O solution.  Finally, the devices were dried on a hotplate.

\begin{figure}[b!]
\centering
\includegraphics[width=0.9\linewidth]{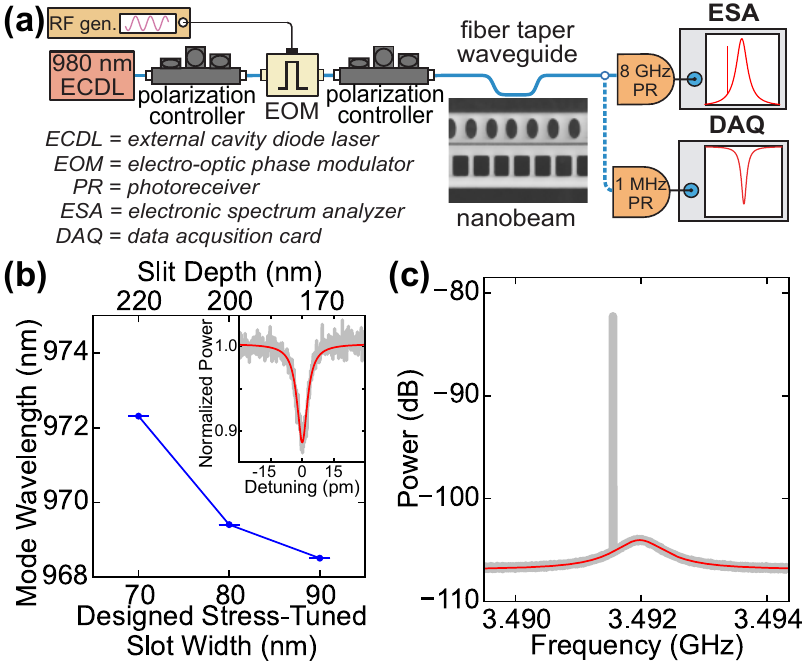}
\vspace{-0.1in}
\caption{(a) Optical modes are detected by swept-wavelength spectroscopy, while mechanical modes are measured when the laser is on the blue-detuned shoulder of the optical mode. For $g_0$ calibration, the laser is phase-modulated.  (b) Optical resonant wavelength of three devices with different stress-tuned slot widths. (inset) Optical spectrum and fit of highest measured $Q_o$ among these devices, having designed gap of 50~nm and $Q_o=\left(1.65 \pm 0.09\right)\times 10^5$~\cite{optErrNote}  (c) Example mechanical spectrum, including phase modulator calibration peak. This power spectral density plot is referenced to a power of 1~mW = 0~dB. Lorentzian fit of thermal noise spectrum is in red.}
\label{fig:Meas}
\end{figure}

A scanning-electron microscope (SEM) image of a released device is shown in Fig.~\ref{fig:StressAndFab}c.  The lithographically-defined slots were between 80~nm and 120~nm, and, with the SEM, we measured stress-tuned slots as small as 24~nm at the center, an aspect ratio of about 10:1 that would be difficult to achieve with lithography alone.  Fig.~\ref{fig:StressAndFab}b graphs the SEM-measured displacements of the beam centers with respect to the stress-tuning slit depths.  The measured trend matches well with the displacements predicted in the FEM simulations.

\section{Demonstration of Slot-Mode Concept}
\label{sec:initmeas}
The experimental setup used to characterized the $\SiN$ slot-mode optomechanical crystals is shown in Fig.~\ref{fig:Meas}a, and was previously described in \cite{ref:Davanco_nanobeam_OMC}.  All measurements were taken at room temperature and pressure.  Devices were characterized with a 980~nm external cavity tunable diode laser, which was coupled evanescently to the devices via a dimpled optical fiber taper waveguide (FTW) with a minimum diameter of $\approx1$~\textmu m.

Among the measured devices, a device with a 50~nm stress-tuned slot had the highest intrinsic optical quality factor $Q_o$ at $\left(1.65 \pm 0.09\right) \times 10^5$ (linewidth of $2.0~\text{GHz}\pm0.1~\text{GHz}$)~\cite{optErrNote}, as shown in Fig.~\ref{fig:Meas}b.  $Q_o$s up to $\approx2.4$ times higher have been demonstrated in $\SiN$ single-nanobeam optomechanical crystals \cite{ref:Grutter_OMC}, but it is expected that the slot mode would have lower $Q_o$s because the geometry has more scattering sites near the optical mode.  In these devices, narrower slots generally resulted in lower $Q_o$s, with 20~nm slot devices having the lowest $Q_o$s around $2.5\times 10^4$.  With further optimization, improving $Q_o$ in smaller slot designs is feasible.

We also used optical characterization to more precisely determine the effect of the stress tuning.  Iterations of devices were made with the same optical and mechanical design but stress-tuning slits of varying depth, so that the only difference among these devices would be the final, stress-tuned slot width.  An example is shown in Fig.~\ref{fig:Meas}b.  Three devices with the same optical and mechanical design show a red shift of the optical resonance as the designed stress-tuned slot width decreases (the stress-tuning slit depth increases).  This trend is expected from simulation (Fig.~\ref{fig:Schem}d), and was consistent in 24 of 27 unique device designs, indicating that varying this stress-tuning slit depth is a reliable technique for tuning the slot width.

\begin{figure}[htbp]
\centering
\includegraphics[width=\linewidth]{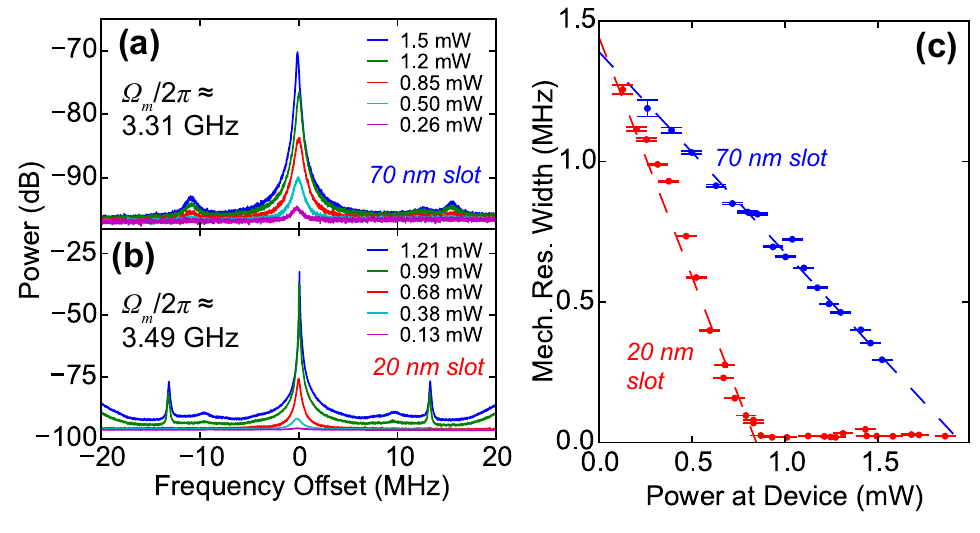}
\vspace{-0.2in}
\caption{(a) Mechanical spectra at different input optical powers ($P_{\text{in}}$) for a device with a designed stress-tuned slot width of 70~nm, intrinsic $Q_o = \left(3.7\pm 0.1\right)\times 10^4$, intrinsic $Q_m = 2380\pm 90$~\cite{mechErrNote}, and $\Omega_m/(2\pi)\approx 3.31$~GHz. (b) Mechanical spectra at different $P_{\text{in}}$ for a device with 20~nm designed stress-tuned slot width, intrinsic $Q_o = \left(3.2\pm 0.1\right)\times 10^4$, intrinsic $Q_m = 2400\pm300$, and $\Omega_m/(2\pi)\approx 3.49$~GHz. (c) Measured $\gamma_{m,\text{eff}}/(2\pi)$ of the devices from (a) (blue) and (b) (red). Error bars represent the uncertainty in the fit of the mechanical spectra to a Lorentzian. Dashed lines show weighted linear fits of the subthreshold $\gamma_{m,\text{eff}}/(2\pi)$. The power spectral density plots in (a) and (b) are referenced to a power of 1~mW = 0~dB.}
\label{fig:DevComp}
\end{figure}

For mechanical mode spectroscopy, the signal was detected with a high-bandwidth (8~GHz) photoreceiver, the output of which was sent to a real-time electronic spectrum analyzer.  Optomechanical characterization required longer-term stability of the coupling, so the FTW was positioned a few hundred nanometers to the side of the device and affixed via van der Waals forces to nearby protruding parts of the $\SiN$ film.  The coupling distance was chosen for a transmission minimum around 70~\%.  The blue detuning of the laser further increased the measurement stability by enabling access to the thermally self-stable regime~\cite{carmon2004dynamical} so that the laser did not have to be externally locked to the cavity.

We used a calibration signal from a phase modulator to measure $g_0$ in a few devices~\cite{ref:Gorodetsky_Kippenberg_OM, ref:KCB_gaas_OM_optica}, as shown in Fig.~\ref{fig:Meas}c, where the phase modulator calibration tone is shown along with the thermal noise spectrum of the 3.49~GHz mechanical breathing mode (quality factor $Q_m\approx3900$).  For a device with a designed, stress-tuned slot width of 60~nm, we measured $g_0/2\pi=184~\text{kHz}\pm2~\text{kHz}$, where the uncertainty comes from the uncertainty in the thermal noise spectrum fit and the measurement of the phase modulator $V_\pi=2.78~\text{V}\pm0.01~\text{V}$ (Sec.~\ref{ref:vpi}). This value matches well with the FEM-simulated $g_0$ values (Fig.~\ref{fig:Schem}f).  Another device, having a designed, stress-tuned slot width of 20~nm, had a phase-modulator-calibrated $g_0/2\pi=317~\text{kHz}\pm3~\text{kHz}$, which also aligns with FEM simulations and confirms the significant improvement in coupling achieved by producing narrow slots.  We note that such slot mode geometries are of particular importance for materials such as $\SiN$, the low refractive index (vs. Si or GaAs) of which limits the achievable coupling strength in single nanobeam geometries.

In addition, with the laser blue-detuned ($\Delta > 0$), optomechanical back-action coherently amplifies the mechanical mode, increasing the detected amplitude while decreasing the effective mechanical linewidth $\gamma_{m,\text{eff}}$. Assuming only optomechanical damping changes with input power, the effective linewidth is related to the optical power at the coupling point to the device $P_{\text{in}}$ as follows, where $\kappa$ is the intrinsic optical loss rate, $\kappa_{\text{ex}}$ is the external coupling rate, $\omega_o$ is the optical resonant frequency, and $\Omega_m$ is the intrinsic mechanical frequency \cite{ref:Kippenberg_Vahala_OE, ref:Eichenfield_thesis}:
\begin{subequations}
\begin{alignat}{3}
    \gamma_{m,\text{eff}} &= \gamma_m + \frac{g_0^2}{\omega_o \hbar}\frac{\kappa_{\text{ex}} P_{\text{in}}}{\Delta^2 + \left(\kappa/2\right)^2}&&\left(\frac{\kappa/2}{\left(\Delta + \Omega_m\right)^2 + \left(\kappa/2\right)^2} \right.\nonumber\\
    & &&\quad \left. {} - \frac{\kappa/2}{\left(\Delta - \Omega_m\right)^2 + \left(\kappa/2\right)^2}\right)\\
    &= \rlap{$\gamma_m + g_0^2 \; \text{S}(\kappa, \kappa_\text{ex}, \omega_o, \Delta, \Omega_m) \; P_{\text{in}}$}
\end{alignat}
\end{subequations}
Thus, the effective mechanical linewidth should change linearly with respect to optical power, with the intercept indicating the intrinsic mechanical linewidth $\gamma_m$ and the slope proportional to $g_0^2$.  For a blue-detuned laser, this slope is negative, and when the optomechanical amplification cancels out $\gamma_m$, the device reaches the regime of regenerative self-oscillation.  The $P_{\text{in}}$ at which this occurs is the threshold power.

We use this relationship to determine the intrinsic $Q_m$ of these devices by looking at the detected mechanical spectrum with respect to power.  To compensate for the cavity's power-dependent thermo-optic shift, for each input power, we adjust the laser wavelength to the optimal detuning value, which corresponds to the point at which the mechanical peak is maximized.  We then linearly fit the subthreshold $\gamma_{m,\text{eff}}$ with respect to $P_{\text{in}}$ to find $\gamma_m$. This same procedure is used to compare devices with similar optical and mechanical parameters; in this case, the slope is an indicator of the relative effective $g_0$.

\begin{figure}[b!]
\centering
\includegraphics[width=0.85\linewidth]{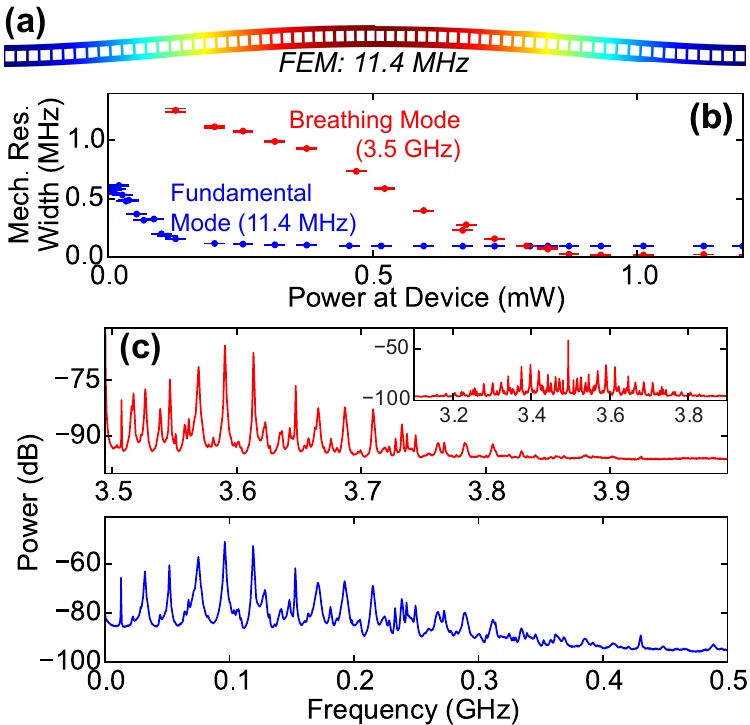}
\caption{(a) FEM simulation of the fundamental lateral flexural beam mode.  (b) Measured 3~dB linewidth of the fundamental flexural beam mode (blue) and the breathing mode (red) as a function of $P_{\text{in}}$.  Error bars represent the uncertainty in the fit of the mechanical spectra to a Lorentzian.  The fundamental mode self-oscillates at $P_{\text{in}}\approx150$~\textmu W, while the breathing mode self-oscillates at $P_{\text{in}}\approx900$~\textmu W.  (c) Sidebands on the breathing mode (red) and the spectrum of harmonics of the lower-frequency flexural beam modes (blue) line up, indicating a mixing between the two. (inset) The full, double-sided spectrum around the self-oscillating breathing mode. All power spectral density plots in (c) are referenced to a power of $1~\text{mW}=0~\text{dB}$.}
\label{fig:FundMode}
\end{figure}

\begin{figure*}[htbp]
\centering
\includegraphics[width=0.85\linewidth]{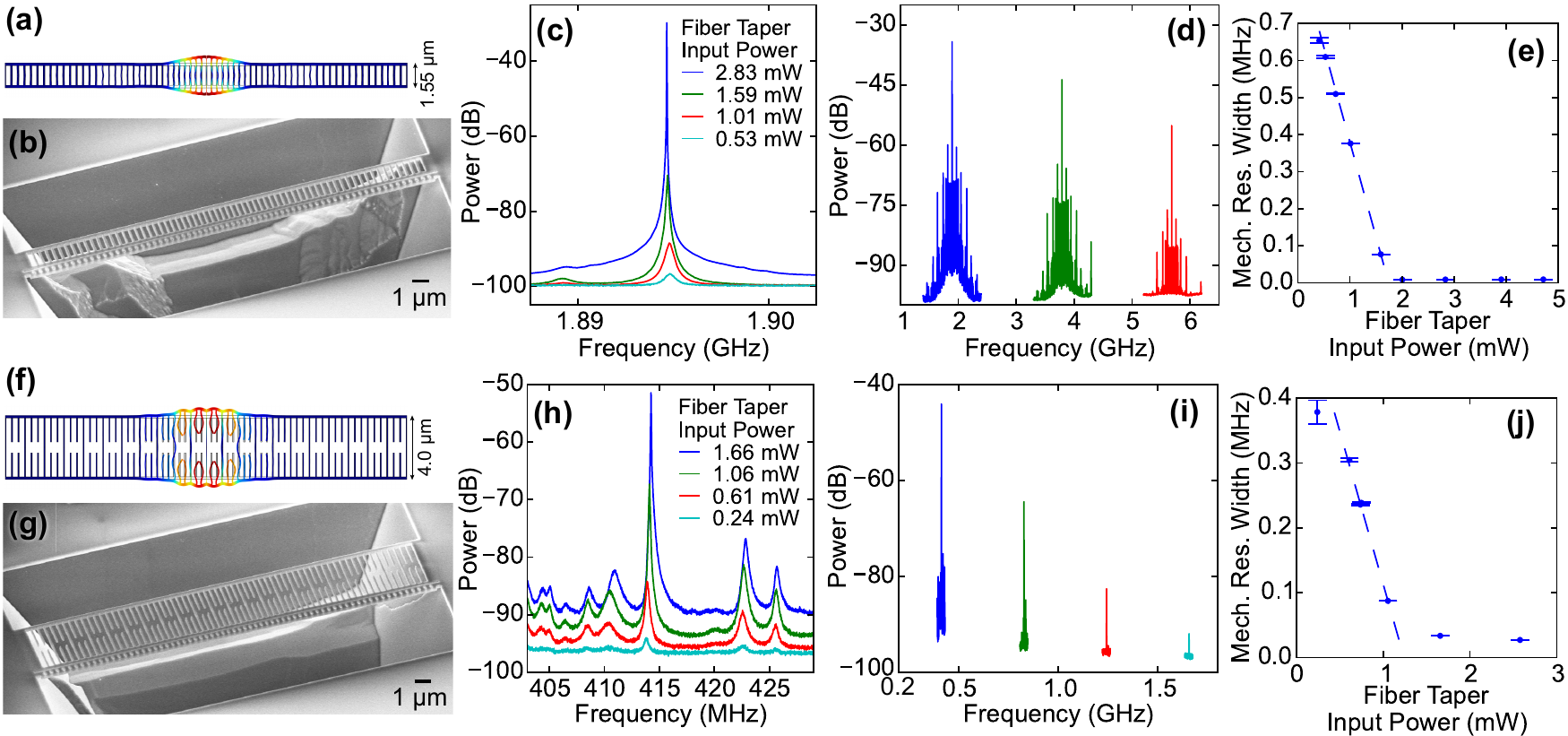}
\caption{(a) FEM simulation of the 1.8~GHz band mechanical breathing mode of the 1.55~$\mum$ wide mechanical beam.  (b) SEM image of a fabricated 1.8~GHz band device.  (c) Detected mechanical spectra at different FTW input optical powers. (d) At a FTW input optical power of 4.7~mW, harmonics of the 1.895~GHz mechanical mode are visible.  (e) Measured $\gamma_{m,\text{eff}}/(2\pi)$ of the 1.895~GHz mechanical mode. Error bars represent the uncertainty in the fit of the mechanical spectra to a Lorentzian. The dashed line shows the weighted linear fit of the subthreshold $\gamma_{m,\text{eff}}/(2\pi)$.  (f) FEM simulation of the 400~MHz band mechanical breathing mode of the 4~$\mum$ wide mechanical beam. (g) SEM image of a fabricated 400~MHz band device.  (h) Mechanical spectra measured at different FTW input optical powers.  (i) At a FTW input optical power of $\approx2.6$~mW, harmonics of the 414~MHz mechanical mode are visible.  (j) Measured $\gamma_{m,\text{eff}}/(2\pi)$ of the 414~MHz mechanical mode. Error bars represent the uncertainty in the fit of the mechanical spectra to a Lorentzian. The dashed line shows the weighted linear fit of the subthreshold $\gamma_{m,\text{eff}}/(2\pi)$. The power spectral density plots in (c), (d), (h) and (i) are referenced to a power of 1~mW = 0~dB.}
\label{fig:LFMFData}
\end{figure*}

Fig.~\ref{fig:DevComp}a and b show measurements of two such devices with similar optical and mechanical performance but different stress tuning.  One device, which had a designed stress-tuned slot width of 70~nm (stress-tuning slit depth of 220~nm), had an intrinsic $Q_o=\left(3.7\pm 0.1\right)\times 10^4$ and an intrinsic $Q_m=2380\pm90$~\cite{mechErrNote}.  The data corresponding to this device are shown in Fig.~\ref{fig:DevComp}a and the blue data in Fig.~\ref{fig:DevComp}c.  The other device had a designed stress-tuned slot width of 20~nm (stress-tuning slit depth of 295~nm), an intrinsic $Q_o=\left(3.2\pm 0.1\right)\times 10^4$, and an intrinsic $Q_m=2400\pm300$. The data corresponding to this narrower-slot device are shown in Fig.~\ref{fig:DevComp}b and the red data in Fig.~\ref{fig:DevComp}c.

Comparing measurements of these two devices, the mechanical mode in the 20~nm slot device is more amplified than in the 70~nm slot device. For $P_{\text{in}}\approx1.2$~mW, the detected mechanical peak in the 20~nm slot device is $\approx65$~dB above the noise floor, while the 70~nm slot device's mechanical peak is only $\approx19$~dB above the noise floor. The measurements of the optomechanical narrowing of the effective mechanical linewidth (Fig.~\ref{fig:DevComp}c), show that the slope of the line for the narrower-slot device (red) is much steeper than for the wider-slot device (blue).  Because they have similar optical and mechanical $Q$s, this suggests that the device with the 20~nm slot has a higher effective $g_0$.  It is also noteworthy that the narrower stress-tuned slot enhances the back-action enough that it reaches self-oscillation above a threshold power of 900~\textmu W.

Among all the devices measured, devices with more aggressive stress tuning (narrower slots) generally had steeper linewidth-narrowing slopes, implying higher effective $g_0$s, as expected from simulation (Fig.~\ref{fig:Schem}f) and confirmed by the aforementioned phase modulator calibration measurements.  The most aggressively tuned devices, with slots designed to be 20~nm, had high enough effective $g_0$s that all but one of them reached the threshold for self-oscillation within the power range of the laser.  This indicates that narrowing the slot via stress-tuning is an effective way to enhance the optomechanical coupling in slot-mode optomechanical crystal devices.

We note that the measured threshold powers are much lower, and the mechanical-linewidth-narrowing slopes much steeper, than would be expected with the $g_0$ values obtained from the phase modulator calibration method.  This suggests other factors in the system are contributing to the effective optomechanical back-action.  These could include DC optical gradient forces acting to pull the beams closer~\cite{ref:eichenfield1} and interaction of the breathing mode with the oscillating flexural beam modes.

In particular, although we designed these devices for optimal coupling to the mechanical breathing mode, and focused our measurements on characterizing it, there are other mechanical modes that couple to the optical mode.  Defects in the fabricated device give rise to additional breathing-type mechanical modes~\cite{ref:eichenfield2}, but the most well-coupled modes tend to be the lateral flexural beam modes.  An FEM simulation of the fundamental lateral flexural beam mode (11.4~MHz) of the mechanical beam is shown in Fig.~\ref{fig:FundMode}a.  Because it is well-coupled to the optical mode and has a much lower frequency than the mechanical breathing mode, its threshold power for self-oscillation is very low; we measure it to be at $P_{\text{in}}\approx150$~\textmu W. We also note that, upon detection of the optical signal modulated by self-oscillating breathing and flexural modes we observed mixing tones as sidebands of the breathing mode, as shown in Fig.~\ref{fig:FundMode}c.

\section{Flexible Mechanical Resonator Design}
\label{sec:LFMF}
Separating the mechanical and optical modes into two beams in the slot-mode architecture adds flexibility in designing these modes compared to a single nanobeam.  By modifying the design of the mechanical beam, a wide range of mechanical frequencies can be accessed without significantly affecting the optical mode. To that end, we demonstrate lower-frequency designs around 1.8~GHz and 400~MHz. Implementing highly-localized breathing modes in various RF bands (here, the IEEE-defined UHF and L) broadens the potential application space. One way to change the mechanical frequency is simply to change the full width of the mechanical beam while keeping the same lattice variation.


For the 1.8~GHz band design, shown in Fig.~\ref{fig:LFMFData}a, the mechanical beam width was increased from 700~nm to 1.55~\textmu m.  Measurements of a fabricated device (Fig.~\ref{fig:LFMFData}b) having a designed, stress-tuned slot width of 80~nm are shown in Fig.~\ref{fig:LFMFData}c-e.  The measured intrinsic $Q_o=\left(1.01\pm0.03\right)\times10^5$, and the measured intrinsic $Q_m = 2130\pm50$, as derived from the weighted linear fit shown in Fig.~\ref{fig:LFMFData}e. These values are comparable to the 3.4~GHz band devices.  The optomechanical coupling of this device was strong enough that we observed self-oscillation for laser powers above $\approx2$~mW ($P_{\text{in}}$ is $\approx20$~\% of the laser power at the FTW input).  Above threshold, we also observed harmonics on the breathing mode (Fig.~\ref{fig:LFMFData}d).  These arise from nonlinear modulation of the optical field due to the Lorentzian optical mode shape, as reported in other systems~\cite{hossein2006characterization, rocheleau2013enhancement, tallur2011monolithic}.

\begin{figure*}[htbp]
\centering
\includegraphics[width=\linewidth]{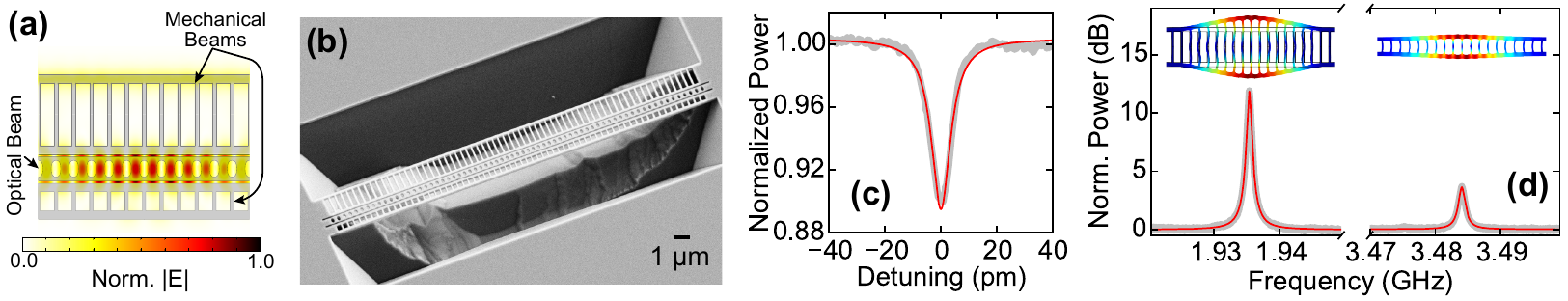}
\caption{(a) FEM simulation of the optical mode of an M-O-M device designed for coupling to 3.4~GHz band (bottom beam) and 1.8~GHz band (top beam) mechanical breathing modes.  The optical mode is in both slots simultaneously.  (b) SEM image of a fabricated M-O-M device. (c) Optical spectrum of M-O-M device.  Measurement is in gray, and the Lorentzian fit is in red.  Measured intrinsic $Q_o=\left(1.26\pm 0.02\right)\times 10^5$ (d) Both mechanical modes measured simultaneously, FTW input power $\approx 3$~mW. Data are in gray and Lorentzian fits are in red.  At this optical input power, the 1.93~GHz mode has effective $Q_m=3175\pm2$, and the 3.484~GHz mode has effective $Q_m=3350\pm10$, where uncertainty comes from 95~\% confidence interval of fit. (insets) FEM eigenmode simulations of corresponding mechanical breathing modes.}
\label{fig:MOMData}
\end{figure*}

For the 400~MHz design, the mechanical beam width was increased to 4~\textmu m.  At this width, the mechanical mode is not well-confined for the same lattice parameters, but the ``ribs'' still contribute to the optical confinement.  Thus, we kept the ribs to maintain high $Q_o$, but increased the effective mechanical lattice constants by ``breaking'' two-thirds of the ribs, as shown in Fig.~\ref{fig:LFMFData}f. Measurements of a fabricated device (Fig.~\ref{fig:LFMFData}g) having a slot width of 80~nm are shown in Fig.~\ref{fig:LFMFData}h-j.  The measured intrinsic $Q_o=\left(1.02\pm0.02\right)\times10^5$, and the measured intrinsic $Q_m=800\pm300$, as derived from the weighted linear fit shown in Fig.~\ref{fig:LFMFData}j. The $Q_o$ is comparable to that of the 3.4~GHz band devices, indicating that the ``broken-rib'' geometry minimally perturbs the optical mode. The $Q_m$, however, is much lower than in the 3.4~GHz band devices.  This is likely due to an increase in air damping with decreased frequency~\cite{chandorkar2008limits} and an increase in anchor loss from the effective two-thirds decrease in the number of lattice periods in the mirror region of the mechanical beam.  As with the other devices in this work, the mechanical spectra (Fig.~\ref{fig:LFMFData}h) include other, less-well-coupled peaks that correspond to additional breathing-type mechanical modes from defects in the fabricated device~\cite{ref:eichenfield2} or harmonics of lower-frequency flexural modes. The optomechanical coupling of the 414~MHz breathing mode of this device was strong enough that it reached self-oscillation for laser powers above $\approx1.2$~mW.  As with the 1.8~GHz band device, we observed harmonics on the breathing mode above threshold (Fig.~\ref{fig:LFMFData}i).

\section{Multimode Optomechanical Devices}
\label{sec:TNBs}
In addition to increasing flexibility in the available mechanical frequencies, the slot-mode device architecture enables new functionality in that it is straightforward to add another separate optical and/or mechanical mode.  In this work, we demonstrate two cases: a single optical mode simultaneously coupled to two different mechanical beams (``M-O-M'') and a single mechanical mode coupled to two different optical modes (``O-M-O'').  M-O-M devices have a variety of possible applications, with theoretical proposals including mechanical mode entanglement and phonon pair generation~\cite{massel2012multimode, wang2013reservoir} and ground-state laser cooling of an unresolved-sideband mechanical resonator~\cite{ojanen2014ground}.  Moreover, recent progress has been made in studying M-O-M devices in other platforms experimentally, including recent investigations of Bogoliubov mechanical modes~\cite{dong2014optomechanically}, as well as systems showing synchronization of mechanical resonators via a travelling optical mode~\cite{zhang2012synchronization, bagheri2013photonic}.  An O-M-O slot-mode device provides a new platform for optical frequency conversion, as proposed in Ref.~\cite{ref:Davanco_OMC}.  Unlike in previous demonstrations of optomechanically-enabled optical frequency conversion~\cite{ref:Hill_Painter_WLC_Nat_Comm, ref:Liu_yuxiang_wlc, ref:Dong_Wang_Science_dark_mode}, the O-M-O device enables quasi-independent optical mode selection and independent optimization of the coupling into each optical mode.

\begin{figure*}[htbp]
\centering
\includegraphics[width=\linewidth]{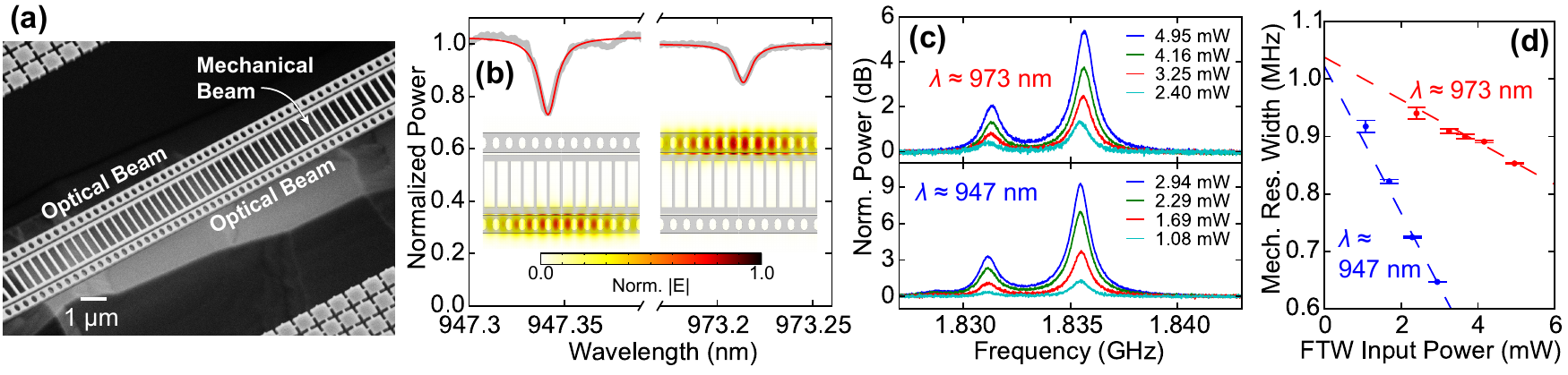}
\caption{(a) SEM image of fabricated O-M-O device.  (b) Separately-measured optical spectra of O-M-O device.  Data are in gray, and Lorentzian fits are in red.  The 947.34~nm mode (``bottom'' beam) has intrinsic $Q_o=\left(1.1\pm 0.1\right)\times 10^5$, and the 973.21~nm mode (``top'' beam) has intrinsic $Q_o=\left(1.05\pm 0.02\right)\times 10^5$. (insets) FEM simulations of the optical slot modes associated with bottom and top optical beams. (c) Mechanical spectra measured at different FTW input optical powers. Top spectra were acquired while optically coupled to the top beam, and bottom spectra were acquired while optically coupled to the bottom beam. (d) $\gamma_{m,\text{eff}}/(2\pi)$ as measured via the top optical mode (red) and the bottom optical mode (blue) with respect to FTW input power. Dashed lines show weighted linear fits of  $\gamma_{m,\text{eff}}/(2\pi)$. Error bars represent the uncertainty in the fit of the mechanical spectra to a Lorentzian.}
\label{fig:OMOData}
\end{figure*}

In the example M-O-M device of this work, we surround an optical beam with a 1.8~GHz band mechanical beam (Sec.~\ref{sec:LFMF}) and a 3.4~GHz band mechanical beam, with 80~nm slots between the beams (Fig.~\ref{fig:MOMData}b). The resultant optical mode is concentrated in both slots simultaneously (Fig.~\ref{fig:MOMData}a).  We couple to the optical mode by hovering the FTW a few hundred nanometers above the optical beam.  The measured intrinsic $Q_o=\left(1.26\pm 0.02\right)\times 10^5$ (Fig.~\ref{fig:MOMData}c), and we simultaneously detect modulation from both the 1.8~GHz band and 3.4~GHz band mechanical modes (Fig.~\ref{fig:MOMData}d).  For the same input optical power, the detected 1.8~GHz band mode has a larger amplitude than the 3.4~GHz band mode primarily because a lower frequency mode has a larger thermal noise motional amplitude for the same temperature.  With the optical quality factor in excess of $10^5$ (linewidth $\approx 2.4$~GHz), this device is in the range of sideband-resolved operation for both the 3.4~GHz and 1.8~GHz band modes, suggesting that this device is a candidate for high-frequency Bogoliubov mechanical mode studies. We have also measured M-O-M devices with 1.8~GHz band and 400~MHz band mechanical breathing modes coupled to the same 981.85~nm optical mode (Sec.~\ref{ref:MOM2}).

The O-M-O device demonstrated here comprises a mechanical beam with a 1.8~GHz band mechanical breathing mode coupled to an optical beam on each side, with 80~nm slots between the beams (Fig.~\ref{fig:OMOData}a).  The top nanobeam was made slightly wider, resulting in a red-shifted optical resonance.  The top and bottom optical modes were characterized separately by repositioning the FTW, and the measured optical spectra are shown in Fig.~\ref{fig:OMOData}b.  The top mode had a measured intrinsic $Q_o=\left(1.05\pm 0.02\right)\times 10^5$ at 973.21~nm, and the bottom mode had a measured intrinsic $Q_o=\left(1.1\pm 0.1\right)\times 10^5$ at 947.34~nm.

The mechanical breathing mode at $\approx1.835$~GHz was detected when coupled both to the top and to the bottom optical modes (Fig.~\ref{fig:OMOData}c).  (There is another, less-well-coupled peak at $\approx 1.831$~GHz that corresponds to either an additional breathing-type mechanical mode from defects in the fabricated device~\cite{ref:eichenfield2} or a harmonic of a lower-frequency flexural mode.)  We also measured the effective mechanical linewidth as a function of power for both optical modes (Fig.~\ref{fig:OMOData}d).  A weighted linear fit of these measurements indicates that the intrinsic $Q_m$ as measured via each optical mode is in good agreement: from the top mode, $Q_m=1800\pm100$, and from the bottom mode, $Q_m=1800\pm200$. The resonant frequency also matches, as shown in Fig.~\ref{fig:OMOData}c, implying that these two optical modes are in fact coupled to the same mechanical mode.  The difference in detected mechanical peak heights and the difference in the slopes of mechanical linewidth with respect to optical power stem from the fact that the bottom optical mode at 947~nm couples more strongly to the mechanical mode.

\section{Discussion}
\label{sec:Disc}
We have demonstrated slot-mode optomechanical devices in which the mechanical breathing mode of a patterned nanobeam is coupled to an optical mode that is laterally confined by a second patterned nanobeam and resides within the slot between the two beams. Along with large optomechanical coupling rates in excess of 300~kHz (as measured via phase-modulator calibration) enabled in part by narrow slot widths that can be achieved by taking advantage of the tensile film stress in $\SiN$, this platform allows for flexible design of the optical and mechanical modes, with mechanical beams tailored to support breathing modes ranging from 400~MHz to 3.5~GHz. Moreover, this geometry can naturally be extended to multimode systems; we have shown triple-nanobeam devices with two different mechanical modes coupled to a single optical mode, as well as a triple-nanobeam device in which two different optical modes are coupled to a single mechanical mode.

Future work will focus on the use of these multimode geometries in applications such as optical wavelength conversion and Bogoliubov mechanical mode formation for phonon pair generation.  Though some of the current devices are already weakly in the sideband-resolved limit ($\kappa/2\pi\approx2~\text{GHz}<\Omega_m/2\pi\approx3.4~\text{GHz}$), additional improvements in $Q_o$ would enable sideband resolution for all of the mechanical frequencies studied.  Finally, the implementation of on-chip waveguides will likely be necessary to achieve long-term, stable coupling to multimoded systems.

\section*{Funding Information}
DARPA (MESO); National Research Council Research Associateship Program


\begin{thebibliography}{10}
\bibitem{mechErrNote}
{Throughout this work, uncertainty in intrinsic $Q_m$ comes from the 95~\%
  confidence interval of the intercept of the weighted linear fit of
  $\gamma_{m,\text{eff}}$ versus optical power. Weighting is from the
  uncertainty in the Lorentzian fits of each mechanical spectrum.}
\end{thebibliography}

\newpage
\appendix
\onecolumngrid
\vskip68pt%
\begin{center} {{\bf \large Supplementary Material}}\end{center}
\vskip11pt
{%
\begin{adjustwidth}{24pt}{24pt}
\rule{\linewidth}{.4pt}
\vskip12pt%
{This section provides supplementary information for this work.  It describes the procedure for determining the phase modulator $V_\pi$, presents data from an additional triple-nanobeam device, and explores how higher-order optical slot modes couple to various mechanical modes.}
\vskip8pt
\noindent\rule{\linewidth}{.4pt}
\end{adjustwidth}
}%
\vskip25pt

\twocolumngrid

\setcounter{figure}{0}
\makeatletter
\renewcommand{\thefigure}{S\@arabic\c@figure}

\setcounter{equation}{0}
\makeatletter
\renewcommand{\theequation}{S\@arabic\c@equation}

\section{Measuring Phase Modulator $V_\pi$}
\label{ref:vpi}
In Sec.~\ref{sec:initmeas}, we use the phase modulator calibration method~\cite{ref:gorodetskysuppl, ref:balramsuppl} to measure the optomechanical coupling $g_0$ of slot-mode optomechanical crystals.  The $V_\pi$ of the electro-optic phase modulator must be accurately known in order to do this calibration.

\begin{figure}[htbp]
\centering
\includegraphics[width=0.75\linewidth]{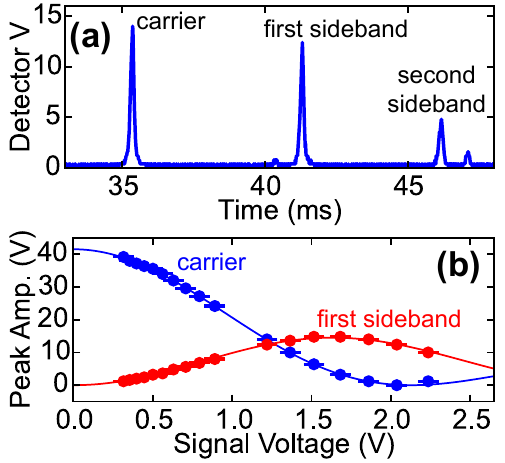}
\caption{(a) An example of the output of the scanning Fabry-P\'{e}rot interferometer for a phase-modulated optical input signal, as read by an oscilloscope. Spectrum shows the carrier peak and the first and second sidebands.  (b) Carrier (blue) and first sideband (red) peak heights with respect to RF signal voltage applied to the phase modulator. Points are the measured values, with error bars indicating the voltage resolution of the oscilloscope. Lines are the fits of the data.}
\label{fig:Vpi}
\end{figure}

To measure $V_\pi$, we send the 980~nm laser signal through the phase modulator, modulated at 3.5~GHz by an RF signal generator, and into a scanning Fabry-P\'{e}rot interferometer.  The detected signal traces out the carrier and phase-modulator-induced sidebands in the optical signal, and we view them on an oscilloscope, as shown in Fig.~\ref{fig:Vpi}a.

Changing the power applied by the RF signal generator to the phase modulator changes the magnitude of the carrier and sidebands.  The RF power $P_{\text{RF}}$ is related to the signal voltage $V_{\text{sig}}=\sqrt{2ZP_{\text{RF}}}$, where the phase modulator input impedance $Z=50~\Omega$.  Knowing this, we can graph the peak magnitudes with respect to $V_{\text{sig}}$, as shown in Fig~\ref{fig:Vpi}b.  For a phase modulator, the carrier peak magnitude should follow the curve $A \left(J_0\left(\pi V_{\text{sig}}/V_\pi\right)\right)^2$, and the first sideband magnitude should follow the curve $A \left(J_1\left(\pi V_{\text{sig}}/V_\pi\right)\right)^2$, where $A$ scales the amplitude of the Bessel functions of the first kind $J_0$ and $J_1$.  We fit data from the carrier and first sideband to these functions in Fig.~\ref{fig:Vpi}b, and both fits result in $V_\pi=2.78\pm0.01$~V, where the uncertainty comes from the fit and is one standard deviation. This value corresponds well with the vendor-specified value for the phase modulator.

\begin{figure*}[htbp]
\centering
\includegraphics[width=\linewidth]{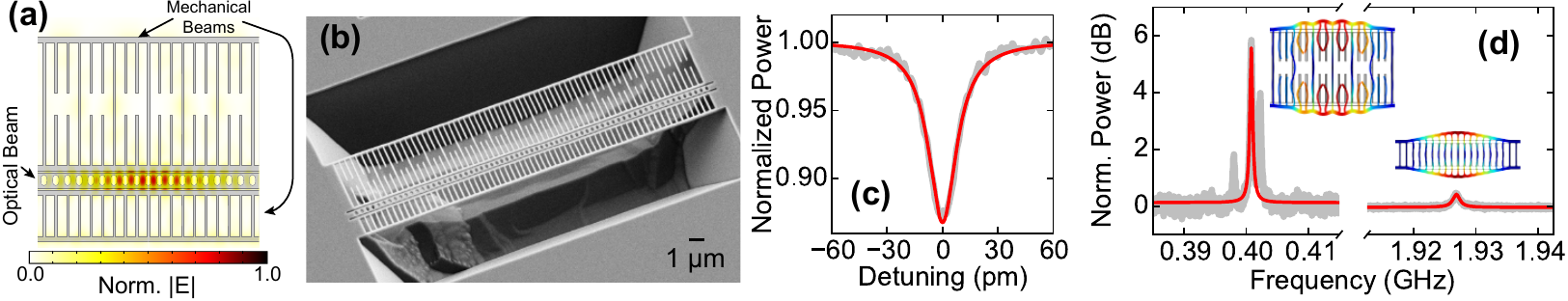}
\vspace{-0.1in}
\caption{(a) FEM simulation of the optical mode of an M-O-M device designed for coupling to 1.8~GHz band (bottom beam) and 400~MHz band (top beam) mechanical breathing modes.  The optical mode is in both slots simultaneously.  (b) SEM image of a fabricated M-O-M device. (c) Optical spectrum of M-O-M device.  Measurement is in gray, and the Lorentzian fit is in red.  Measured intrinsic $Q_o=\left(5.80\pm 0.06\right)\times 10^4$ (d) Both mechanical modes measured simultaneously, FTW input power $\approx 2$~mW. Data are in gray and Lorentzian fits are in red.  At this optical input power, 400~MHz mode has effective $Q_m=1030\pm20$, and 1.927~GHz mode has effective $Q_m=1450\pm20$, where uncertainty comes from 95~\% confidence interval of fit. (insets) FEM eigenmode simulations of corresponding mechanical breathing modes.}
\label{fig:MOMLM}
\end{figure*}

\section{Additional M-O-M Device}
\label{ref:MOM2}
In addition to the example M-O-M device in Sec.~\ref{fig:MOMData}, we fabricated and characterized an M-O-M device in which a 400~MHz band mechanical beam and a 1.8~GHz band mechanical beam surround an optical beam, with 80~nm slots between the beams, shown in Fig.~\ref{fig:MOMLM}b. The resultant optical mode is concentrated in both slots simultaneously (Fig.~\ref{fig:MOMLM}a).  We couple to the optical mode by hovering the FTW a few hundred nanometers above the optical beam.  The measured intrinsic $Q_o$ was $\left(5.80\pm 0.06\right)\times 10^4$ (Fig.~\ref{fig:MOMLM}c), and we simultaneously detect modulation of the transmitted optical signal from both the 400~MHz band and 1.8~GHz band mechanical modes (Fig.~\ref{fig:MOMLM}d).  For the same input optical power, the detected 400~MHz band mode has a larger amplitude than the 1.8~GHz band mode primarily because a lower frequency mode has a larger thermal noise motional amplitude for the same temperature.  As with the other devices in this work, the mechanical spectrum (Fig.~\ref{fig:MOMLM}d) includes other, less-well-coupled peaks that correspond to additional breathing-type mechanical modes from defects in the fabricated device~\cite{ref:eichenfield2suppl} or harmonics of lower-frequency flexural modes.

\section{Coupling to Higher-Order Slot Modes}
\label{ref:HOmodes}

\begin{figure}[htbp]
\centering
\includegraphics[width=0.85\linewidth]{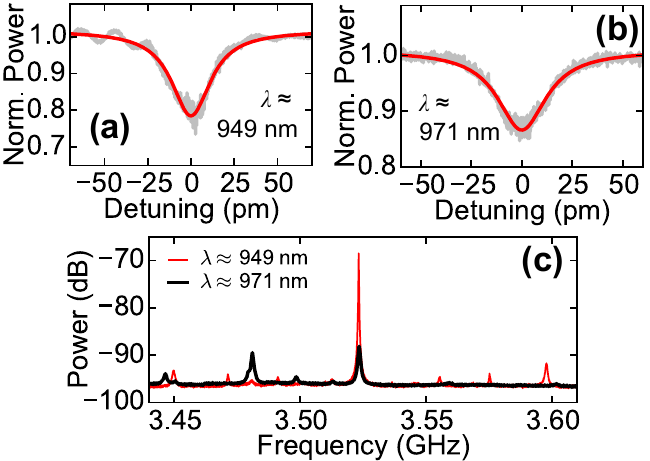}
\vspace{-0.1in}
\caption{(a) Fundamental optical slot mode at $\approx949$~nm with intrinsic $Q_o = \left(3.96\pm0.09\right)\times10^4$. (b) Higher-order optical slot mode in the same device at $\approx971$~nm with intrinsic $Q_o = \left(3.57\pm0.08\right)\times10^4$. (c) Mechanical spectra measured while coupled to the fundamental optical mode (red) and higher-order optical mode (black).  The optical power input to the fiber taper waveguide was $\approx2.2$~mW at the 949~nm mode and $\approx3.5$~mW at the 971~nm mode. This power spectral density plot is referenced to a power of 1~mW = 0~dB.}
\label{fig:HOopt}
\end{figure}

These slot-mode optomechanical crystals confine multiple optical modes in the slot in addition to the fundamental mode for which they were designed.  We observed some of these higher-order modes, as shown in Fig.~\ref{fig:HOopt}a and b.  Because they are distributed more widely along the slot, these modes couple less strongly to the highly-localized breathing mode and more strongly to other mechanical modes in the device, such as higher-order breathing-type mechanical modes arising from fabrication defects~\cite{ref:eichenfield2suppl}.

In the example of Fig.~\ref{fig:HOopt}, a device with a designed, stress-tuned slot width of 20~nm has an optical mode at $\approx949$~nm with an intrinsic optical $Q = \left(3.96\pm0.09\right)\times10^4$ as well as an optical mode at $\approx971$~nm with an intrinsic optical $Q = \left(3.57\pm0.08\right)\times10^4$. When pumped at the 949~nm fundamental optical mode, it self-oscillates at the $\approx3.52$~GHz mechanical breathing mode, and we begin to see sidebands due to mixing with the modulation from the low-frequency flexural beam modes, as described in Sec.~4 of the main text.  However, even with 50~\% more optical power, the mechanical breathing mode does not self-oscillate when pumped at the 971~nm optical mode.  In addition, the mechanical spectrum reveals another peak at $\approx3.48$~GHz with about the same optomechanical coupling to the 971~nm mode as the $\approx3.52$~GHz mode, suggesting that this higher-order optical mode is also coupling to some higher-order, less-well-confined mechanical mode.


\end{document}